\pdfoutput=1

\documentclass[10pt]{iopart}

\usepackage{graphicx}
\usepackage{iopams}  
\usepackage{hyperref}

\newcommand{\allpart}{\pi^{\pm}, K^{\pm}, p, \bar{p}, \phi, \Lambda, \mathrm{and}~\bar{\Lambda}}
\newcommand{\sq}{\sqrt{s_{NN}}}
\newcommand{\vslope}{dv_{1}/dy}

\newcommand{\pT}{p_T}

\begin{document}

\title[Charge-Dependent Directed Flow]{Charge-Dependent Directed Flow in Symmetric Nuclear Collisions}

\author{Vipul Bairathi$^1$ and Kishora Nayak$^{2*}$}
\address{$^1$Instituto de Alta Investigación, Universidad de Tarapacá, Casilla 7D, Arica 1000000, Chile}
\address{$^2$Department of Physics, Panchayat College, Bargarh 768028, Odisha, India}
\ead{$^*$k.nayak1234@gmail.com}
\vspace{10pt}


\begin{abstract}
The directed flow ($v_1$) of identified hadrons ($\pi^{\pm}, K^{\pm}, p, \bar{p}, \phi, \Lambda$, and $\bar{\Lambda}$) is studied in symmetric nuclear collisions (O+O, Cu+Cu, Ru+Ru, Au+Au, and U+U) at $\sqrt{s_{NN}} = 200$ GeV using the string-melting version of a multiphase transport model with improved quark coalescence. The mid-rapidity $v_1$-slope ($dv_1/dy$) and its charge-dependent splitting ($\Delta dv_1/dy$) between particles and anti-particles are investigated as a function of nuclear mass number ($A$) and collision centrality in both low-$p_\mathrm{T}$ (0.2$-$2.0 GeV/$c$) and high-$p_\mathrm{T}$ (2.0$-$5.0 GeV/$c$) regions. At low-$p_\mathrm{T}$, the $v_1$-slope shows weak system-size dependence, while at high-$p_\mathrm{T}$ strong system-size dependence is found and it becomes negative with nuclear mass number, reflecting the hard-soft asymmetry in particle production. The charge-dependent splitting $\Delta dv_1/dy$ reveals a striking baryon-meson dichotomy: baryon pairs ($p-\bar{p}$ and $\Lambda-\bar{\Lambda}$) exhibit significant splitting that grows with system size, whereas meson pairs ($\pi^+-\pi^-$ and $K^+-K^-$) show minimal splitting. 
The effect of final state hadronic interactions on the $v_1$-slope is found to be negligible confirming
that it is primarily generated during the partonic phase and coalescence process. A comparison of the AMPT results with measurements from the STAR experiment at RHIC in Au+Au collisions establishes the transported quark contribution as a baseline for the observed charge-dependent $v_1$ splitting, on top of which electromagnetic field effects must be considered.
\end{abstract}

\vspace{2pc}

%
\maketitle

\ioptwocol

\section{Introduction}
\label{sec:Intro}
Azimuthal anisotropic flow is one of the key observables for characterizing the properties of the medium produced in relativistic heavy-ion collisions~\cite{BRAHMS,phobos,star2005,phenix,heinz_snellings}. Among the various flow coefficients ($v_{n}$), the directed flow ($v_{1}$) represents the first-order harmonic in the Fourier expansion of the azimuthal distribution of final-state particles relative to the reaction plane. This Fourier expansion can be expressed as~\cite{voloshin_zhang,flow6}:
\begin{equation}
	\frac{dN}{d\phi} \propto 1 + \sum_{n=1}^{\infty}2v_n(\pT,y)\cos[n(\phi-\Psi_{RP})],
\end{equation}
where $\pT$, $y$, and $\phi$ denote the transverse momentum, rapidity, and azimuthal angle of the particles in momentum space, respectively, and $\Psi_{RP}$ is the reaction plane angle. In the above Fourier expansion, the sine terms vanish due to the reflection symmetry with respect to $\Psi_{RP}$.

The directed flow, $v_{1} =\langle\cos(\phi-\Psi_{RP})\rangle$, is an odd function of rapidity, $v_{1}^{\mathrm{odd}}(y) = -v_{1}^{\mathrm{odd}}(-y)$, due to the geometry of the collisions. It is sensitive to the early-time dynamics of heavy-ion collisions~\cite{bozek_wyskiel,Snellings2000,Stocker2005}, including the pressure gradients from tilted source geometry and the electromagnetic fields generated by spectator protons~\cite{dflow4,dflow6}. The directed flow of identified hadrons provides unique insights into particle-type dependent responses to these initial conditions, as different hadron species carry distinct information about the underlying quark dynamics, specifically, the contributions from transported quarks (originating from initial-state nucleons) versus produced quarks (created during the collision). The magnitude of directed flow is generally expressed in terms of slope at mid-rapidity, $\vslope|_{y=0}$, and will be referred to as $\vslope$ throughout this paper.

Extensive experimental measurements of $v_{1}$ have been carried out at both the RHIC and the LHC~\cite{exdflow1,exdflow2,exdflow3,exdflow4,exdflow8,exdflow9,exdflow10}. Recently, the STAR experiment at RHIC performed measurements of $\Delta \vslope$ between particles and anti-particles for $\pi$, $K$, and $p$ in various centrality classes~\cite{STAR:2023jdd,star_systsize_pid_v1}. A sign change of $\Delta \vslope$, especially for $p-\bar{p}$ from central to peripheral collisions, is observed, and it is attributed to the presence of a strong electromagnetic (EM) field, which increases from central to peripheral collisions. Recent studies have employed hydrodynamical models with and without electromagnetic fields to interpret the results from the STAR experiment~\cite{sandeep_1,sandeep_2}. 

Transport models, such as the AMPT with string melting (AMPT-SM), are quite successful in describing particle spectra and flow coefficients in heavy-ion collisions~\cite{newCoal,AMPTv1,AMPTv1PLB,our_ampt_epjc}. In a recent study (Ref.~\cite{our_ampt_epjc}), we have employed the new coalescence AMPT-SM model to study system size dependence of $v_{1}$ and its slope for inclusive charged hadrons at $\sq$ = 200 GeV. This dependence is attributed to the hard-soft asymmetry in the particle production profile of the produced medium. 

In this work, we aim to understand the experimental findings on the system size dependence of $v_{1}$ splitting between oppositely charged hadrons. We focused on studying the $\Delta\vslope$ of particle-antiparticle pairs with the same mass, such as $\pi^{+}-\pi^{-}$, $K^{+}-K^{-}$, $p-\bar{p}$, and $\Lambda-\bar{\Lambda}$~\cite{exdflow8,exdflow6}. We report $v_{1}$ of identified hadrons ($\allpart$) for both low- and high-$\pT$ in O+O, Cu+Cu, Ru+Ru, Au+Au, and U+U collisions at $\sq$ = 200 GeV using the same AMPT-SM model. We studied system size dependence of $v_{1}$-slope through $\vslope$ as a function of nuclear mass number and collision centrality. The geometric scaling property is tested by dividing $\vslope$ by $\mathrm{A}^{1/3}$ in both the soft and hard particle production scenarios. Among the hadrons studied, the production of $K^{-}$, $\phi$, $\overline{p}$, and $\overline{\Lambda}$ has contributions from quarks produced during the collision. However, the production of $\pi^{\pm}$, $K^{+}$, $p$, and $\Lambda$ includes contributions from both quarks produced in the collision and those transported from the initial colliding nucleons. Therefore, it is essential to study charge-dependent $v_{1}$ and $\Delta \vslope$ in symmetric nuclear collisions using the AMPT model, which does not explicitly incorporate the electromagnetic field. 

The paper is organized into the following sections. Section~\ref{sec:model} provides a brief description of the AMPT model, including the incorporation of nuclear deformation in the model. Section~\ref{subsec:analysis} presents the analysis technique and results, highlighting the dependence of $v_1$, $\vslope$, and $\Delta\vslope$ between particle and anti-particle on the system size in symmetric heavy-ion collisions. The centrality and mass number dependence of identified hadrons $v_1$ at low- and high-$\pT$ is discussed. The effect of hadronic interaction time on $v_1$-slope is also discussed. Finally, Sec.~\ref{sub:summary} summarizes the findings and provides an outlook for future research.

\section{The AMPT model overview}
\label{sec:model}
The AMPT model primarily consists of four stages: (i) Initial conditions, which are derived from the HIJING model and include the spatial and momentum distributions of minijet partons and excited strings~\cite{Hijing94}; (ii) Parton interactions,  which employ Zhang’s parton cascade model for parton scatterings, including only two-body scatterings with cross sections obtained from perturbative QCD with screening masses~\cite{Zpc98}; (iii) Fragmentation, which involves the conversion of partonic matter to hadronic matter. In the default mode, a Lund string fragmentation model is used to convert strings into hadrons~\cite{Lund94}. In the string melting mode of the AMPT model, a quark coalescence model is used~\cite{AMPTv1}; and (iv) Hadron interactions, where the dynamics of the subsequent hadronic matter are described by a hadronic cascade based relativistic transport model~\cite{Art01}.

In the AMPT model, the distribution of nucleons within the nuclei is modeled by the Wood-Saxon (WS) distribution function, described by the following equation:
\begin{equation}
\rho(r,\theta) = \frac{\rho_{0}}{1+e^{[\lbrace r - R(\theta,\phi)\rbrace/a]}},
\label{ws_eq}
\end{equation}
where $\rho_{0}$ is the normal nuclear density, $r$ and $a$ are the distance from the center of the nucleus and the surface diffuseness parameter, respectively. The parameter characterizing the deformation of the nucleus is denoted by $R(\theta,\phi)$ as, 
\begin{equation} 
R(\theta,\phi) = R_{0}[1 + \beta_{2}Y_{2,0}(\theta,\phi)].
\end{equation} 
Here, $R_{0}$, $\beta_{2}$, and $Y_{l,m}(\theta,\phi)$ are the radius parameter, quadrupole deformation parameter, and spherical harmonics, respectively.

The AMPT-SM model, with an improved version of the coalescence mechanism, is used in this analysis~\cite{newCoal}. A parton-parton interaction cross-section of 1.5 mb is utilized, consistent with the parameters employed in Refs.~\cite{AMPTv1,our_ampt_epjc}. The default hadron cascade time in the AMPT-SM model is $t_{\rm max} = 0.4$ fm/$c$, which effectively disables final-state hadronic interactions. Events are generated for various collision systems, including O+O, Cu+Cu, Ru+Ru, Au+Au, and U+U at $\sqrt{s_{\mathrm {NN}}}$ = 200 GeV using the corresponding WS parameters as outlined in Table~\ref{tab1}. These parameters are based on Refs.~\cite{nucl_para1,nucl_para2,nucl_para3}. 
\begin{table}[!htbp]
\begin{center}
\caption{Wood-Saxon parameters for various nuclei in the AMPT-SM model~\cite{our_ampt_epjc}.}
\label{tab1}
\begin{tabular}{cccccc} 
Parameter & $^{16}_{8}\rm{O}$  & $^{63}_{29}\rm{Cu}$ & $^{96}_{44}\rm{Ru}$ & $^{197}_{79}\rm{Au}$ & $^{238}_{92}\rm{U}$		\\[0.5ex]
\hline  
$R_{0}$     & 2.608 & 4.214 & 5.090 & 6.380 & 6.810 \\[0.2ex]
$a$         & 0.513 & 0.586 & 0.460 & 0.535 & 0.550 \\[0.2ex]
$\beta_{2}$ & 0 	& 0 	& 0.162	& 0 	& 0.280 \\[0.2ex]
\hline
\end{tabular}
\end{center}
\end{table}

\section{Analysis and results}
\label{subsec:analysis}
The directed flow of identified and charged hadrons in Au+Au collisions has been studied at center-of-mass energies ranging from 7.7 to 200 GeV, as reported in Refs.~\cite{newCoal,AMPTv1,Nayak2019,NayakShiLin2024}. These studies use an AMPT model with an improved coalescence mechanism. Additionally, a Coalescence Sum Rule (CSR) involving seven produced hadrons has been tested~\cite{AMPTv1PLB,AMPT_v1_Univ}. In this paper, we report the rapidity-dependent directed flow of identified hadrons at $\sq$ = 200 GeV across different colliding systems, from O+O to U+U, using the same improved version of the AMPT model with string melting mode. The analysis technique followed in this paper is the same as in Ref.~\cite{our_ampt_epjc}, including the procedure for selecting centrality based on charged-particle multiplicity, similar to the experimental analysis.

\subsection{$v_1(y)$ of identified hadrons}
\label{ssec:v1eta}
\begin{figure*}[!htbp]
\centering
\includegraphics[scale=0.8]{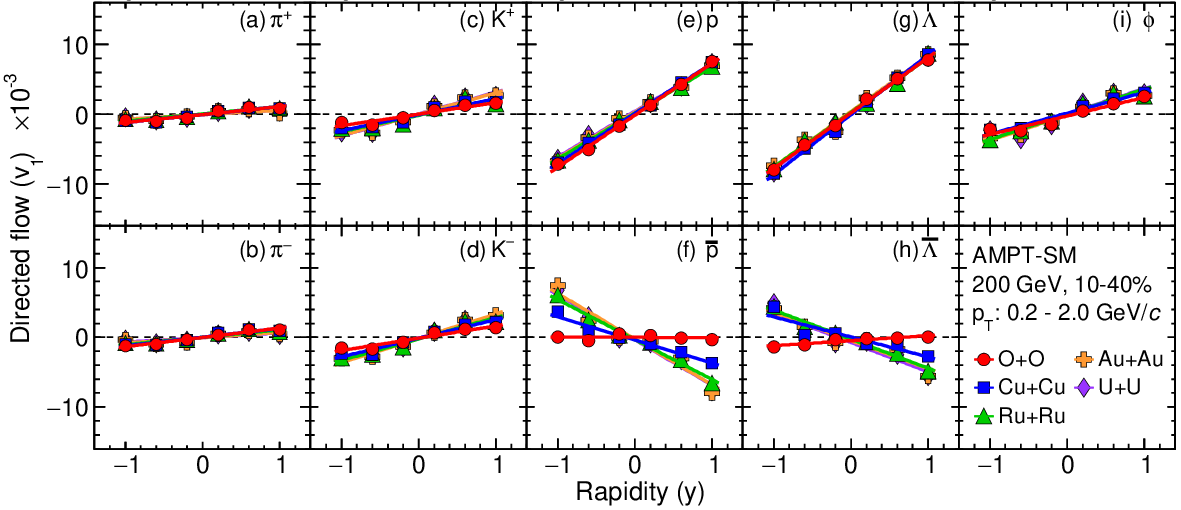}
\caption{$v_1$($y$) of identified hadrons ($\pi^{\pm}$, K$^{\pm}$, $p$, $\bar{p}$, $\Lambda$, $\bar{\Lambda}$, and $\phi$) at low-$\pT$ in different colliding systems (O+O, Cu+Cu, Ru+Ru, Au+Au, and U+U) for 10--40$\%$ centrality at $\sq$ = 200 GeV using the AMPT-SM model.}
\label{fig:v1yLowpT}
\end{figure*}

\begin{figure*}[!htbp]
\centering
\includegraphics[scale=0.8]{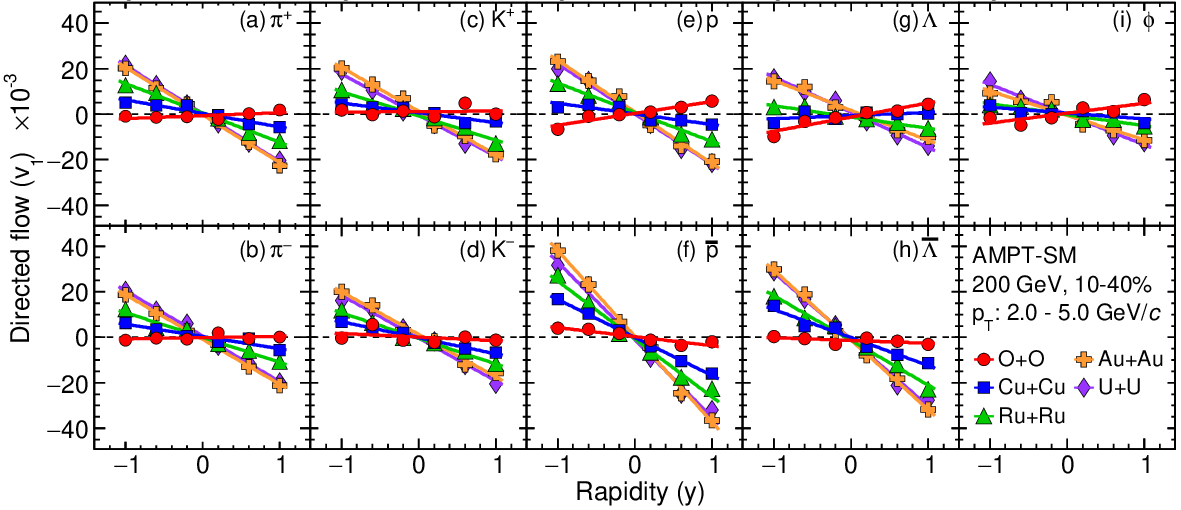}
\caption{$v_1$($y$) of identified hadrons ($\pi^{\pm}$, K$^{\pm}$, $p$, $\bar{p}$, $\Lambda$, $\bar{\Lambda}$, and $\phi$) at high-$\pT$ in different colliding systems (O+O, Cu+Cu, Ru+Ru, Au+Au, and U+U) for 10--40$\%$ centrality at $\sq$ = 200 GeV using the AMPT-SM model.}
\label{fig:v1yHighpT}
\end{figure*}
The rapidity-dependent directed flow $v_1(y)$ for mesons ($\pi^{\pm}$, $K^{\pm}$, $\phi$) and baryons ($p$, $\bar{p}$, $\Lambda$, $\bar{\Lambda}$) in 10--40\% central O+O, Cu+Cu, Ru+Ru, Au+Au, and U+U collisions at $\sq$ = 200 GeV, calculated using the AMPT-SM model, is shown in Fig.~\ref{fig:v1yLowpT} and Fig.~\ref{fig:v1yHighpT}. The results are presented for low-$\pT$ (0.2 $< \pT <$ 2.0 GeV/$c$) and high-$\pT$ (2.0 $< \pT <$ 5.0 GeV/$c$) regions to explore soft (thermal) and hard (jet-associated) particle production mechanisms. The $v_1(y)$ is fitted with a linear polynomial function of the form $v_1(y) = Fy + C$ within the fitting range of $|y| < 1.0$~\cite{flow6}. This allows us to extract the mid-rapidity directed flow slope parameter $F \equiv \vslope|_{y=0}$, which serves as the primary observable for comparing different particle species and collision systems.

The low-$\pT$ charged hadrons, except $\bar{p}$ and $\bar{\Lambda}$, show an increasing $v_1$ trend with $y$ from O+O to U+U collisions at $\sq$ = 200 GeV. In contrast, high-$\pT$ charged hadrons show a decreasing $v_1$ trend with $y$ for all collision systems, except for $p$, $\Lambda$, and $\phi$ in O+O collisions. The $\pT$-dependent behavior and system-size evolution of $v_1(y)$ for inclusive charged hadrons has been examined in Ref.~\cite{our_ampt_epjc}, where the underlying particle production mechanism related to hard-soft asymmetry is detailed. In this work, we explored $\pT$ and system-size dependence of $v_1(y)$ for identified hadrons to understand the interplay between initial-state geometry, parton dynamics, and hadronization mechanisms across different collision systems. The extracted slope parameters and their scaling properties are examined in detail in the following subsections.

\subsection{$v_1$-slope of identified hadrons}
\label{sec:systsize}
\begin{figure*}[!htbp] 
\centering
\includegraphics[scale=0.8]{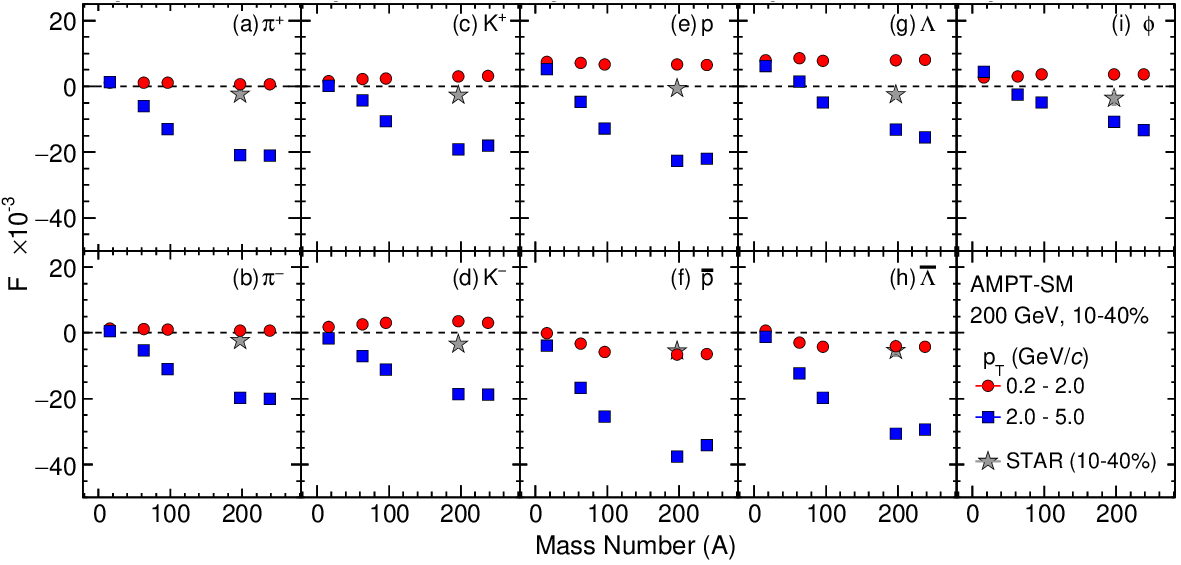}
\caption{$v_1$-slope parameter $F$ versus mass number ($A$) for low- and high-$\pT$ identified hadrons in 10--40$\%$ centrality for O+O, Cu+Cu, Ru+Ru, Au+Au, and U+U collisions at $\sq$ = 200 GeV using the AMPT-SM model.}
\label{fig:dv1dyvsA}
\end{figure*}

Figure~\ref{fig:dv1dyvsA} presents the $v_1$-slope parameter $F$ as a function of mass number (A) for identified hadrons at both low- and high-$\pT$ in 10--40\% central O+O, Cu+Cu, Ru+Ru, Au+Au, and U+U collisions at $\sq$ = 200 GeV using the AMPT-SM model. At low-$\pT$, the $v_1$-slope shows minor or negligible change with varying mass numbers, indicating weak dependence on system size. Both mesons ($\pi^{\pm}$, $K^{\pm}$, and $\phi$) and baryons ($p$ and $\Lambda$) show positive slopes that remain approximately constant across the colliding systems. Interestingly, anti-baryons ($\bar{p}$ and $\bar{\Lambda}$) exhibit negative slopes, consistent with the experimental measurements from the STAR collaboration at RHIC in Au+Au collisions at $\sq$ = 200 GeV~\cite{exdflow3,exdflow6,exdflow8}.

At high-$\pT$, the identified hadron $v_1$-slope shows a significant change with varying mass number, indicating a strong dependence on the size of colliding systems. The magnitude of $v_1$-slope shows a significant difference between low- and high-$\pT$ hadrons. As the system size increases, the slopes become increasingly negative, reaching maximum negative values for the larger systems (Au+Au and U+U). Interestingly, for O+O collisions, $v_1$-slope shows minimal change between low- and high-$\pT$ particles. The observed trends in smaller systems compared to larger systems suggest that the dynamics may be dominated by soft particle production processes, leading to different energy deposition and parton interactions. As a result, the fragmentation processes and subsequent hadronization might yield different particle distributions in smaller systems from those produced in larger collision systems. Therefore, the transition from smaller to larger collision system would be able to shed light on our understanding of the phase transitions and thermalization of the medium produced in these collisions. A comparison of directed flow with experimental data may reveal more about the effects of system size on particle production and the collective behavior of the matter produced in relativistic nucleus-nucleus collisions.

\subsection{Charge-dependent splitting of $v_1$-slope}
\label{ssec:charge}
In the AMPT-SM model, quarks transported from the initial-state nucleons carry a net baryon number and follow different space-time trajectories than produced quarks. Their coalescence into baryons (but not antibaryons) creates an inherent $v_1$ asymmetry. Hence, the system-size dependence arises naturally from the increasing number of transported quarks in larger systems, consistent with enhanced baryon stopping. 

\begin{figure*}[!htbp]
\centering
\includegraphics[scale=0.7]{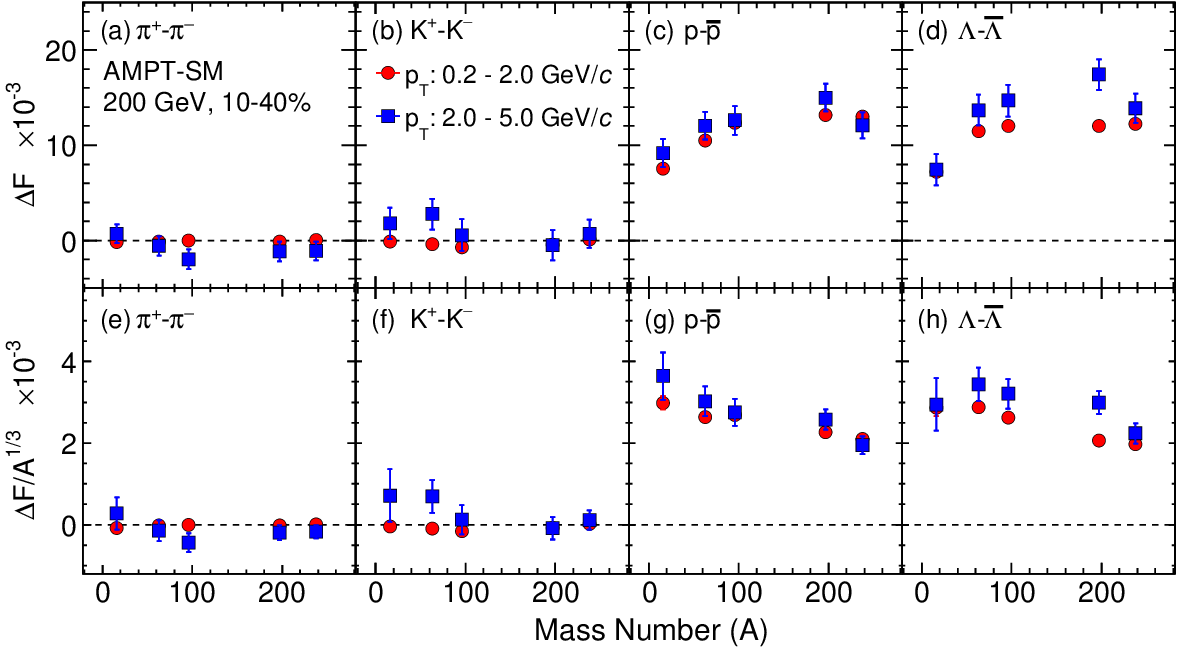}
\caption{Difference in $v_1$-slope parameter ($\Delta F$) between particles and anti-particles (upper panels) and $\Delta F$ scaled by $A^{1/3}$ (lower panels) in 10--40\% central O+O, Cu+Cu, Ru+Ru, Au+Au, and U+U collisions at $\sq$ = 200 GeV using the AMPT-SM model.}
\label{fig:DeltaSlopeMassNum}
\end{figure*}
Figure~\ref{fig:DeltaSlopeMassNum} shows the mass number dependence of the $v_1$-slope splitting between particle-antiparticle pairs ($\Delta F = F_{\rm particle} - F_{\rm antiparticle}$) for low- and high-$\pT$ identified hadrons in 10--40\% central O+O, Cu+Cu, Ru+Ru, Au+Au, and U+U collisions at $\sq$ = 200 GeV. The $\Delta F$ exhibits a striking particle-type dependence: baryon pairs ($p-\bar{p}$ and $\Lambda-\bar{\Lambda}$) show significant splitting that increases with system size and saturates for the largest systems, whereas meson pairs ($\pi^+-\pi^-$ and $K^+-K^-$) show minimal splitting that remains nearly constant across all systems. The $\Lambda-\bar{\Lambda}$ splitting is particularly informative: since $\Lambda$ and $\bar{\Lambda}$ are electrically neutral, their splitting arises solely from baryon number transport in the AMPT model, offering a unique observable for disentangling transported-quark contributions from electromagnetic field effects in future measurements.

The lower panels of Fig.~\ref{fig:DeltaSlopeMassNum} show $\Delta F$ scaled by $A^{1/3}$. Meson pairs exhibit trivial geometric scaling with system size, while baryon pairs show clear deviations---the scaled splitting decreases from O+O to U+U, indicating physics beyond simple geometric scaling. No significant difference in both $\Delta F$ and scaled $\Delta F$ is observed between the low- and high-$\pT$ regions, indicating that the charge-dependent mechanism is imprinted during the early partonic phase and preserved through coalescence. While the $\pT$-differential $\Delta v_1(\pT)$ may exhibit more complex structure~\cite{sandeep_2}, the mid-rapidity $\Delta F$ remains robust across the studied $\pT$ range (0.2 $< \pT <$ 5.0 GeV/$c$).

\subsection{Centrality-dependent splitting of $v_1$-slope}
\label{ssec:cent_dependent}
The STAR Collaboration at RHIC has reported measurements of centrality-dependent $\Delta(\vslope)$ for $\pi^+-\pi^-$, $K^+-K^-$, and $p-\bar{p}$ in Au+Au collisions at $\sq = 200$ GeV~\cite{STAR:2023jdd}. They also conducted measurements in Ru+Ru, Zr+Zr, and Au+Au collisions to explore system size dependence. A sign change in $\Delta(\vslope)$ was observed from central to peripheral collisions across all the studied systems~\cite{star_systsize_pid_v1}. In central collisions, the observed dependence of $\Delta F$ was attributed to enhanced baryon stopping. In peripheral collisions, the Coulomb and Faraday effects of the EM field are suggested to be more dominant and leads to negative values of $\Delta(\vslope)$.

A study based on a hydrodynamic model has been reported in references \cite{sandeep_1,sandeep_2}. This hydrodynamic model incorporates both the EM field and baryon diffusion. The findings indicate a significant splitting ($v_1$) between baryons and anti-baryons, which increases with centrality from central to peripheral collisions in the absence of the EM field and baryon diffusion ($C_B = 0$). When baryon diffusion is included (with $C_B = 1$) but without the EM field, the splitting decreases in peripheral collisions. However, when the EM field is present, the $\Delta(\vslope)$ becomes negative in peripheral collisions, especially when accounting for electrical conductivity and baryon diffusion \cite{sandeep_1,sandeep_2}.

\begin{figure*}[!htbp]
\centering
\includegraphics[scale=0.28]{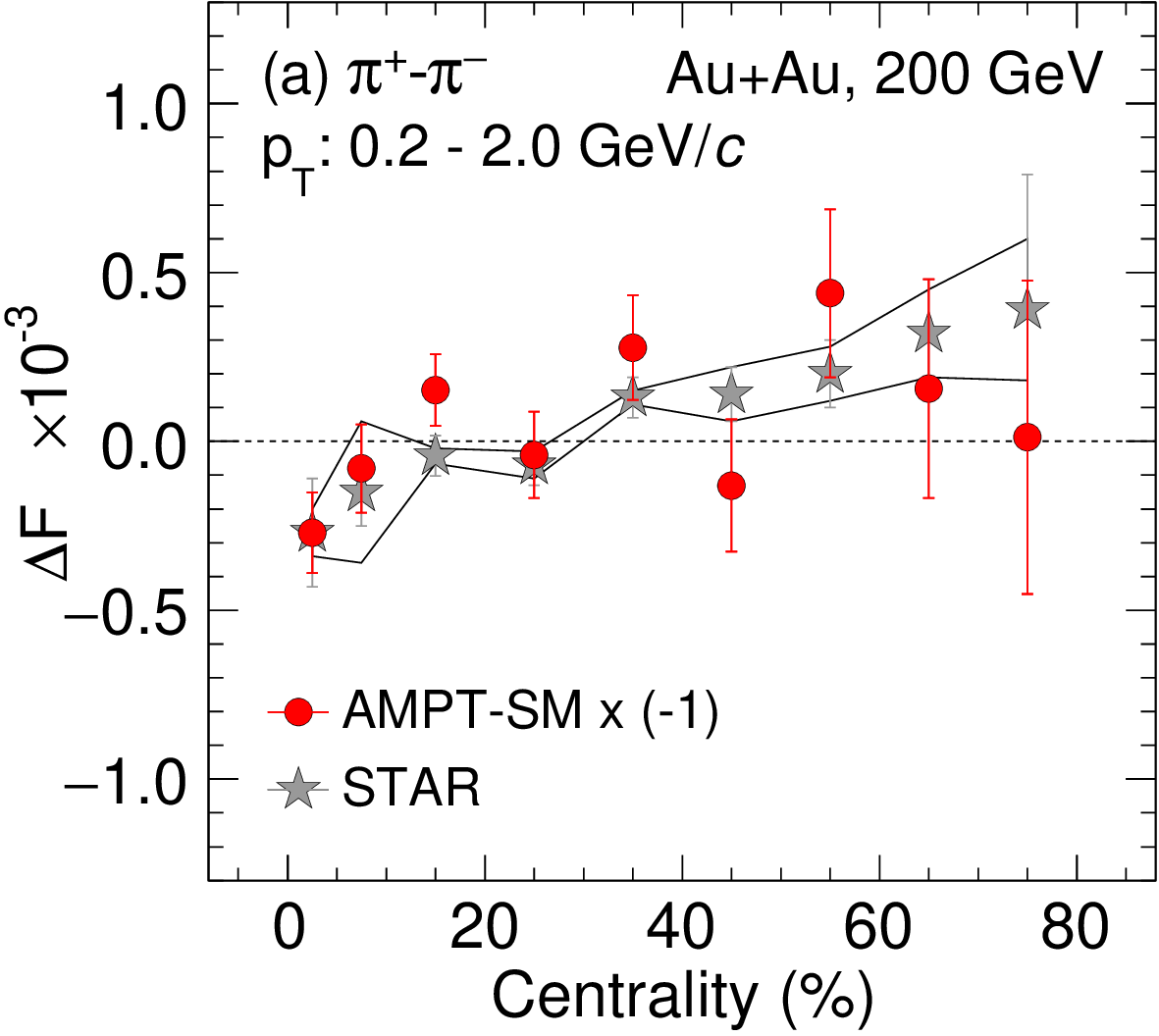} 
\includegraphics[scale=0.28]{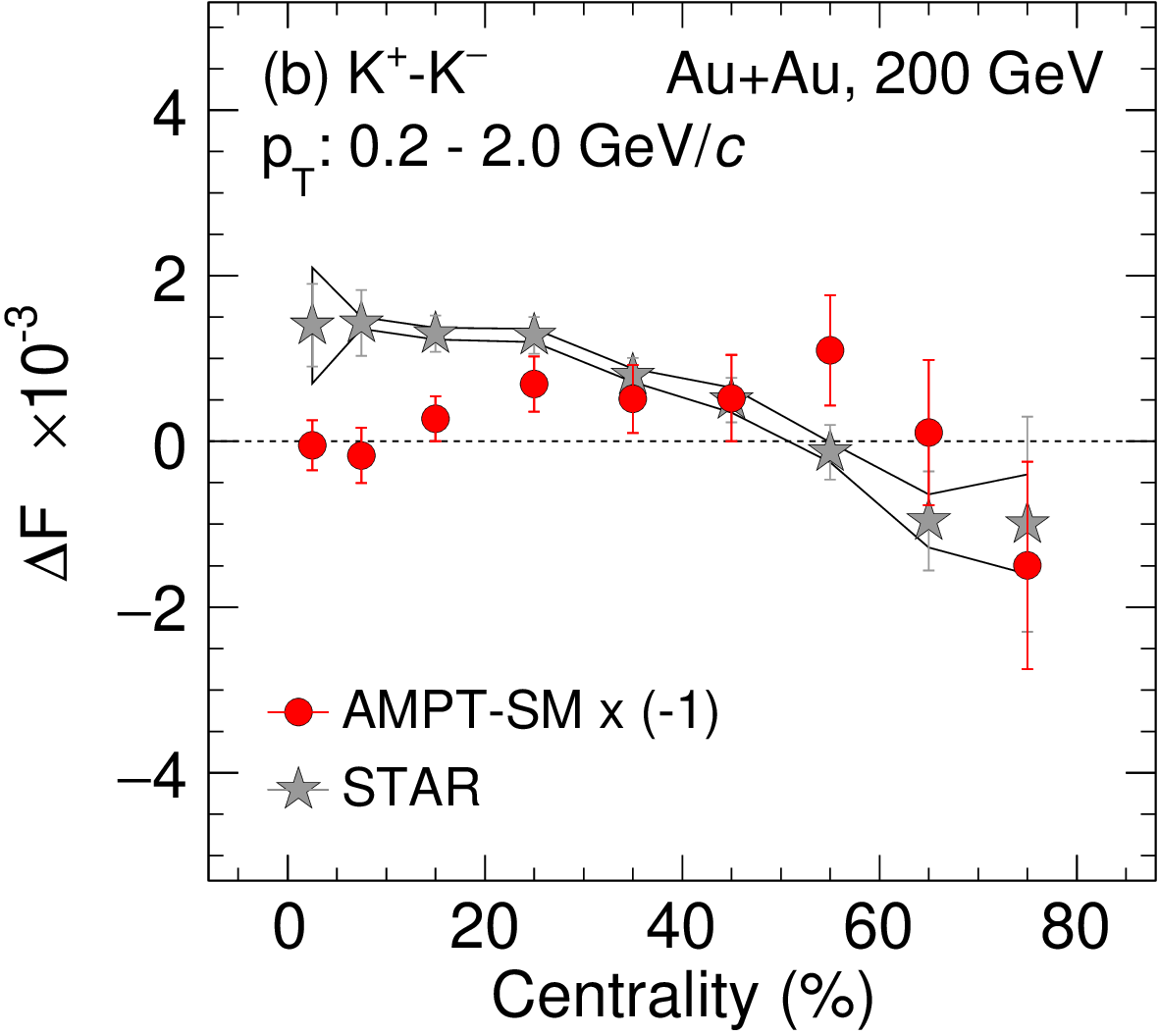}
\includegraphics[scale=0.28]{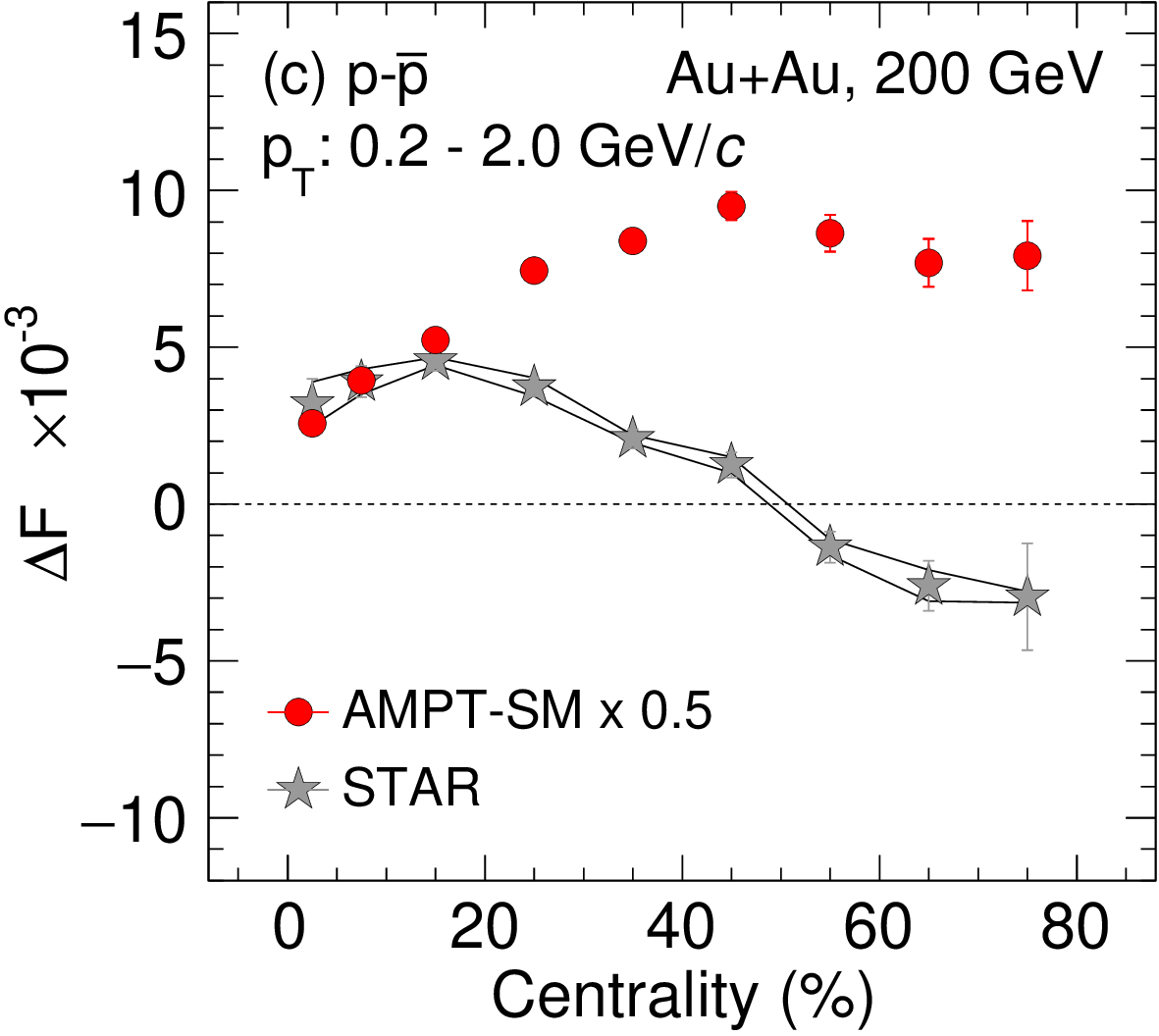}
\caption{Difference in $v_1$-slope parameter ($\Delta F$) as a function of centrality for low-$\pT$ (a) $\pi^+-\pi^-$, (b) $K^+-K^-$, and (c) $p-\bar{p}$ in Au+Au collisions at $\sq =$ 200 GeV from the AMPT-SM model. The results are compared with the measurements fro the STAR experiment at RHIC~\cite{STAR:2023jdd,star_systsize_pid_v1} in Au+Au collisions at $\sq$ = 200 GeV.}
\label{fig:deltav1slopeCent_models}
\end{figure*}
Figure~\ref{fig:deltav1slopeCent_models} presents the centrality dependence of $\Delta F$ for $\pi^+-\pi^-$ , $K^+-K^-$, and $p-\bar{p}$ in Au+Au collisions at $\sq$ = 200 GeV from the AMPT-SM model. The results are compared with the measurements from the STAR experiment~\cite{STAR:2023jdd,star_systsize_pid_v1}. The AMPT-SM results show almost negligible splitting for mesons pairs in different centralities. The AMPT-SM model, which lack explicit EM field implementation, capture the qualitative centrality dependence but underestimate the magnitude of $\Delta F$ for $p-\bar{p}$. The comparisons establish that the AMPT-SM provides a baseline from transported quark dynamics, on top of which EM field effects must be superimposed for quantitative agreement with data.

\begin{figure*}[!htbp]
\centering
\includegraphics[scale=0.85]{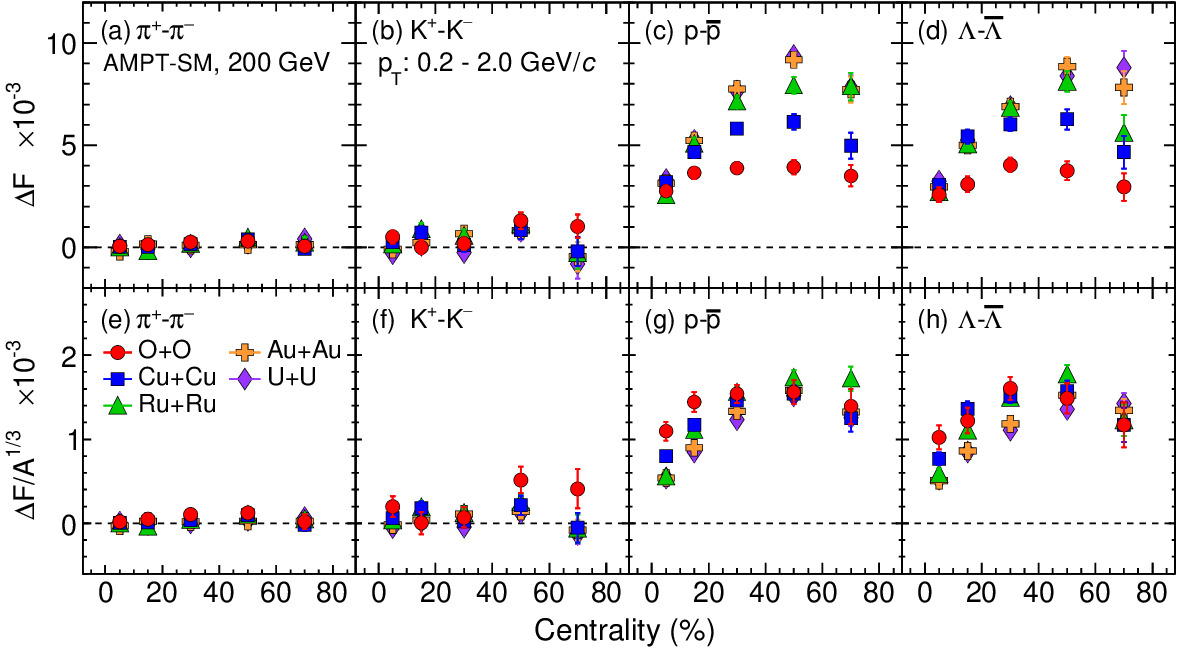}
\caption{Centrality dependence of $\Delta F$ (upper panels) and $\Delta F/A^{1/3}$ (lower panels) for low-$\pT$ identified hadrons in O+O, Cu+Cu, Ru+Ru, Au+Au, and U+U collisions at $\sq$ = 200 GeV using the AMPT-SM model.}
\label{fig:deltav1slopeCent_lowpT}
\end{figure*}
Figure~\ref{fig:deltav1slopeCent_lowpT} shows the centrality dependence of $\Delta F$ and $\Delta F/A^{1/3}$ for low-$\pT$ identified hadrons across all collision systems at $\sq$ = 200 GeV. Meson pairs maintain nearly constant $\Delta F$ from central to peripheral collisions, while baryon pairs exhibit increasing trends from central to mid-central collisions and then decreases toward peripheral collisions. The scaled $\Delta F$ shows similar centrality dependence and approximate scaling for all particle species confirming that the splitting in the AMPT-SM model is primarily governed by geometric factors. 

\subsection{Effect of hadronic interactions}
\label{ssec:hadint}
\begin{figure*}[!htbp]
\centering
\includegraphics[scale=0.85]{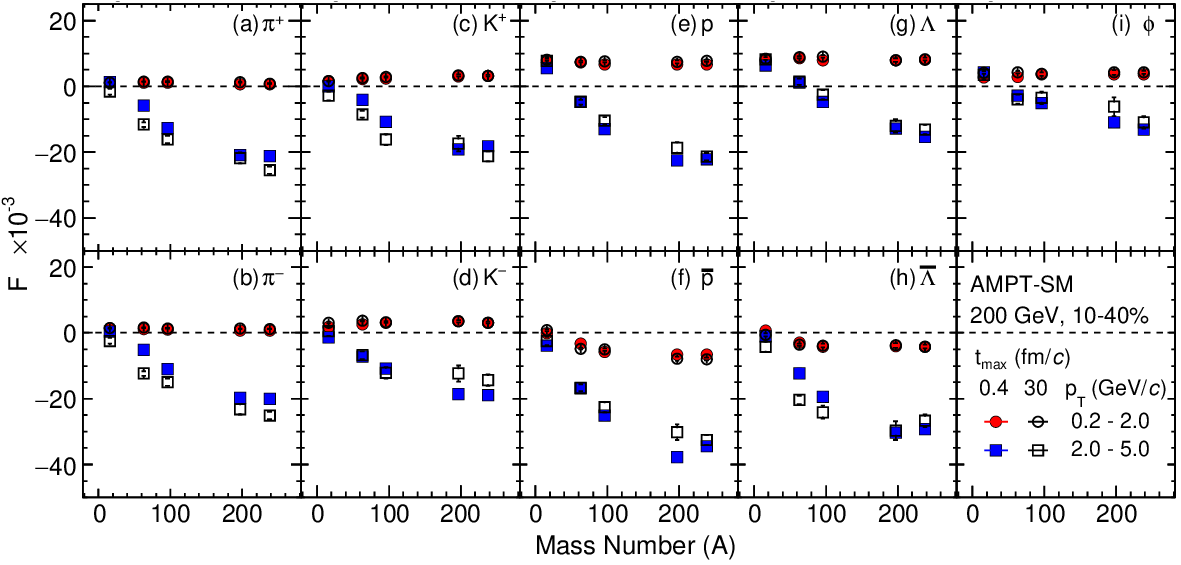}
\caption{The $v_1$-slope parameter $F$ as a function of mass number ($A$) for identified hadrons in 10--40$\%$ central O+O, Cu+Cu, Ru+Ru, Au+Au, and U+U collisions at $\sq$ = 200 GeV using the AMPT-SM model, compared between the default hadronic cascading time of $t_{\rm max} = 0.4$ fm/$c$ (solid symbols) and $t_{\rm max} = 30$ fm/$c$ (open symbols).}
\label{fig:hadint}
\end{figure*}

The results presented in the preceding subsections were obtained with the default hadron cascade time $t_{\rm max} = 0.4$ fm/$c$ in the AMPT-SM model, which effectively disables hadronic rescattering after hadronization. To assess the sensitivity of the $v_1$-slope to late-stage hadronic interactions, the analysis is performed with $t_{\rm max} = 30$ fm/$c$, which enables the full hadronic interactions via the ART model~\cite{Art01}. Figure~\ref{fig:hadint} compares $F$ as a function of mass number between the two hadronic cascading scenarios  in 10--40\% central collisions across all systems at $\sq$ = 200 GeV. The $v_1$-slope for all particle species shows negligible sensitivity to hadronic rescattering time within uncertainty. The particle-type and system-size dependence of the $F$ are preserved, confirming that the directed flow slope is primarily established during the partonic phase and coalescence process, and is not significantly modified by subsequent hadronic interactions.

\section{Summary}
\label{sub:summary}
A comprehensive study has been conducted on the directed flow ($v_1$) and its slope at mid-rapidity ($\vslope$) for identified hadrons ($\pi^{\pm}$, $K^{\pm}$, $\phi$, $p$, $\bar{p}$, $\Lambda$, $\bar{\Lambda}$) in various symmetric colliding systems (O+O, Cu+Cu, Ru+Ru, Au+Au, and U+U) at $\sqrt{s_{\rm NN}} = 200$ GeV using the new coalescence AMPT-SM model. The results are presented for both low-$\pT$ (0.2--2.0 GeV/$c$) and high-$\pT$ (2.0--5.0 GeV/$c$) to explore the role of soft and hard particle production mechanisms. The observed positive $v_1$-slope ($F > 0$) for low-$\pT$ identified hadrons, except anti-baryons ($\bar{p}$, $\bar{\Lambda}$), suggests that soft particles move along with the bulk medium. In contrast, high-$\pT$ hadrons flow in the opposite direction to the bulk medium, resulting in a negative $v_1$-slope ($F < 0$) for all colliding systems except O+O. In the smallest system (O+O), the medium does not appear to generate sufficient opacity to oppose the motion of high-$\pT$ hadrons. In larger systems, the opposite sign of $F$ between low- and high-$\pT$ hadrons reflects the hard-soft asymmetry in the particle production profile, consistent with the findings from our earlier study on inclusive charged hadrons~\cite{our_ampt_epjc}.

The charge-dependent splitting of the $v_1$-slope ($\Delta F = F_{\rm particle} - F_{\rm antiparticle}$) reveals a striking baryon-meson dichotomy. Meson pairs ($\pi^+-\pi^-$, $K^+-K^-$) exhibit minimal splitting that remains nearly constant across all system sizes, consistent with coalescence of produced quarks. In contrast, baryon pairs ($p-\bar{p}$, $\Lambda-\bar{\Lambda}$) show substantial splitting that increases with mass number and saturates for the largest systems. No significant difference in $\Delta F$ is observed between the low- and high-$\pT$ regions, suggesting that the charge-dependent mechanism is imprinted during the early partonic phase and preserved through coalescence. This splitting originates from the asymmetry between transported quarks and produced quarks in the AMPT model and it represents analog of baryon inhomogeneity in the hydrodynamic model calculations~\cite{sandeep_1,sandeep_2}. The $\Lambda-\bar{\Lambda}$ splitting serves as a particularly unique observable, since the $\Lambda$ and $\bar{\Lambda}$ are electrically neutral and their splitting arises solely from baryon number transport, offering a means to disentangle the transported-quark contribution from electromagnetic field effects for the future experimental measurements.

A centrality dependence of $\Delta F$ is observed for baryon pairs, which increases from central to mid-central collisions
and then decreases toward peripheral collisions, while meson pairs remain nearly constant. After $A^{1/3}$ scaling, the splitting for all particle species exhibits nearly independent centrality profile, demonstrating that the charge-dependent splitting in the AMPT-SM model is primarily governed by geometric factors.

A comparison with the STAR measurements~\cite{STAR:2023jdd,star_systsize_pid_v1} in Au+Au collisions shows that the AMPT-SM model qualitatively captures the baryon-meson splitting hierarchy but underestimates the absolute magnitude, particularly for baryons in peripheral collisions where the electromagnetic field effects are expected to be strongest. The AMPT-SM results therefore establish a baseline contribution from baryon transport dynamics, on top of which electromagnetic field effects must be superimposed for quantitative agreement with data. Furthermore, the effect of hadronic rescattering on $F$ is found to be negligible for all particle species, confirming that the directed flow slope is established at the partonic level before hadronization. Especially, the predictions reported here for small systems (O+O and Cu+Cu) provide benchmarks for upcoming experimental measurements and can help isolate the baseline baryon-transport contribution to the directed flow splitting, which is essential for constraining electromagnetic field effects.
 
\section*{Acknowledgments}
\label{acknowledgement}
KN is supported by OSHEC, Department of Higher Education, Government of Odisha, Index No. 23EM/PH/124 under MRIP 2023. The authors thank Prof. B. Mohanty for providing computational facilities at NISER, India. The authors also thank Prof. Z.-W. Lin for the new coalescence AMPT model.

\section*{References}
\bibliographystyle{unsrt}
\bibliography{Reference_PIDv1}

\end{document}